\documentclass[oneside,a4paper,11pt]{article}

\usepackage{biometrika}
\usepackage{amsfonts}
\usepackage{graphicx}
\usepackage{graphics}
\usepackage{epsfig}
\usepackage[round]{natbib}
\usepackage{amsmath}
\usepackage{amsthm}
\usepackage{amsbsy}
\usepackage{amssymb}

\def\N{N}
\def\Un{\mbox{Un}}

\def\IGa{\mbox{{\footnotesize IG}}}

\def\Be{\mbox{Be}}

\def\kmax{\mbox{{$\max\{k\}$}}}
\def\kkmax{\mbox{{\tiny {$\max\{k\}$}}}}

\def\kmaxp{\mbox{{$\max\{k(i,j)\}$}}}

\def\kkkmaxp{\mbox{{\small {$\max\{k(i,j)\}$}}}}

\def\Section{\S}

\def\al{\footnotesize \rm{al}}
\def\d{\footnotesize \rm{d}}

\def\pr{\hbox{pr}}

\def\eqdef{{\equiv}}

\def\qcon{\mbox{$\tilde{c}_i$}}

\newcommand{\comment}[1]{}

\begin{document}

\begin{center}
{\Large \bf
Retrospective Markov chain Monte Carlo methods  for Dirichlet
  process hierarchical model}
\bigskip

{By OMIROS PAPASPILIOPOULOS {\footnotesize AND} GARETH O. ROBERTS}

\medskip
{\em
Department of Statistics,
Warwick University, Coventry, CV4 7AL, U.K.}

{O.Papaspiliopoulos@warwick.ac.uk, \ \ \ \ \ \ \ \
Gareth.O.Roberts@warwick.ac.uk}

\bigskip
{SUMMARY}
\end{center}

Inference for  Dirichlet  process hierarchical models is
typically performed using Markov chain Monte Carlo methods, which
can be roughly categorised into marginal and conditional
methods.
The former integrate out analytically the
infinite-dimensional component of the hierarchical model and
sample from the marginal distribution of the remaining
variables using the Gibbs sampler. Conditional methods impute
the  Dirichlet process and
update it as a component of the Gibbs sampler. Since this requires
imputation of an infinite-dimensional process, implementation of
the conditional method has relied on finite approximations. In
this paper we show how to avoid such approximations by
designing two novel Markov chain Monte Carlo algorithms which
sample from the exact posterior distribution of quantities of
interest. The approximations are avoided by the new technique of
retrospective sampling. We also
show how the algorithms can obtain samples from functionals of the
Dirichlet process.  The marginal and the conditional methods are
compared and a careful simulation study is included, which
involves a non-conjugate model, different datasets and prior
specifications.



\noindent {\em Some keywords:}
Exact simulation; Mixture models; Label switching;
Retrospective sampling; Stick-breaking
prior

\section{Introduction}
\label{intro}

Dirichlet process hierarchical models, also known as Dirichlet
mixture models, are now standard in semiparametric inference;
applications include density estimation
\citep{mull:erka:west:1996}, survival analysis
\citep{gelf:kott:2003}, semi-parametric analysis of variance
\citep{mull:etal:2005}, cluster analysis and partition modelling
\citep{quin:igle:2003}, meta-analysis \citep{burr:doss:2005} and
machine learning \citep{teh:2006}. The hierarchical structure is as
follows. Let $f(y \mid z,\lambda)$ be a parametric density with
parameters $z$ and $\lambda$, let $H_\Theta$ be a distribution indexed
by some parameters $\Theta$ and let $\Be(\alpha,\beta)$ denote the
beta distribution with parameters $\alpha,\beta$. Then we have
\begin{eqnarray}
Y_i \mid   (Z,K,\lambda) & \sim & f(Y_i \mid Z_{K_i},\lambda),~i=1,\ldots,n \nonumber \\
K_i \mid p & \sim & \sum_{j=1}^{\infty}p_j \delta_j(\cdot)
\nonumber \\
Z_j \mid \Theta & \sim & H_{\Theta},~j=1,2,\ldots  \label{hier-cond}\\
p_1 = V_1, & & p_j=(1-V_1)(1-V_2)\cdots (1-V_{j-1})V_j,~j \geq 2\, \nonumber \\
V_j & \sim & \Be(1,\alpha) \nonumber\,.
\end{eqnarray}
Here
$V=(V_1,V_2,\ldots)$ and $Z=(Z_1,Z_2,\ldots)$ are vectors of
independent variables and are independent of each other, the
$K_i$s are independent given $p=(p_1,p_2,\ldots)$, and the $Y_i$s
are independent given $Z$ and $K=(K_1,K_2,\ldots,K_n)$. Therefore,
(\ref{hier-cond}) defines a mixture model in which $f(y \mid
z,\lambda)$ is mixed with respect to the discrete random
probability measure $P({d} z)$, where
\begin{equation}
P(\cdot)= \sum_{j=1}^{\infty} p_j \delta_{Z_j}(\cdot),
\label{series}
\end{equation}
and $\delta_x(\cdot)$ denotes the Dirac delta measure centred at
$x$. Note that, the discreteness of $P$ implies a distributional
clustering of the data. The last three lines in the hierarchy
identify $P$ with the Dirichlet process with base measure
$H_{\Theta}$ and concentration parameter $\alpha$; this
representation is due to \cite{seth:1994}.  For the construction
of the Dirichlet process prior see
Ferguson (1973,1974),
\nocite{ferg:1973} \nocite{ferg:1974} for properties of Dirichlet
mixture models see for example \nocite{lo:1984,ferg:1983,anto:1974}
Lo (1984), Ferguson (1983), and Antoniak (1974),
and for a recent article on modelling with
Dirichlet processes see \cite{green:rich:2001}.
Using more general beta distributions in the last line gives rise to the
rich class of stick-breaking random measures and the general hierarchical
framework introduced by
\cite{ishw:jame:2001}. We have consciously been vague
about the spaces on which $Y_i$ and $Z_j$ live; it should be
obvious from the construction that a great deal of flexibility is
allowed.

We refer to $Z_j$ and $p_j$ as the parameters and the weight
respectively of the $j$th component in the mixture, and to $K_i$ as
the allocation variable, which determines to which component the
$i$th data point is allocated. Note that the prior on the
component weights $p_j$ imposes a weak identifiability on the
mixture components, in the sense that $E(p_j) \geq E(p_l)$
for any $j \geq l$. Throughout the paper we use the nomenclature
`cluster' to refer to a mixture component with at least one datum
allocated to it.
 The hyperparameters
$\alpha,\Theta$ and $\lambda$ will be consider fixed until
\Section \ref{sec:hyper}. We will concentrate on inference for the
allocation variables, the component parameters and weights, and
$P$ itself.

Inference for Dirichlet mixture models  has been made feasible by
Gibbs sampling techniques which have been developed since the seminal
work in an unpublished 1988 Yale University Ph. D. thesis by M. Escobar.
Alternative Monte Carlo schemes do
exist and  include the sequential samplers of \cite{liu:1996} and
\cite{ishw:jame:2003}, and referred to as the { blocked Gibbs sampler} in
those papers, the particle filter of \cite{fear:2004} and
the reversible jump method of \cite{green:rich:2001} and
\cite{jain:neal:2004}. We concentrate on Gibbs
sampling and related componentwise updating algorithms.

Broadly speaking, there are two possible Gibbs sampling
strategies, corresponding to two different data augmentation
schemes. The marginal method exploits convenient mathematical
properties of the Dirichlet process and integrates out
analytically the random probabilities $p$ from the hierarchy.
Then, using a Gibbs sampler, it obtains samples from the posterior
distribution of the $K_i$'s and the parameters and the weights of
the clusters. Integrating out $p$ induces prior dependence among
the $K_i$'s, and makes the labels of the clusters unidentifiable.
This approach is easily carried out for conjugate models, where
further analytic integrations are possible; (\ref{hier-cond}) is
said to be conjugate when $H_\Theta({d} z)$ and $f(y \mid
z,\lambda)$ form a conjugate pair for $z$. Implementation of the
marginal method for non-conjugate models is considerably more
complicated. \cite{mace:mull:1998} and \cite{neal:2000} are the
state-of-the-art implementations in this context.

The conditional method works with the  augmentation scheme
described in (\ref{hier-cond}). It consists of the imputation of
 $p$ and $Z$, and subsequent Gibbs sampling from the joint posterior
distribution of $(K,p,Z)$. The conditional independence structure
created with the imputation of $(p,Z)$ assists the simultaneous
updating of large subsets of the variables. The conditional method
was introduced in \cite{ishw:zarep:2000}, it was further developed
in \cite{ishw:jame:2001, ishw:jame:2003} and it has two
considerable advantages over the marginal method. First, it does
not rely on being able to integrate out analytically components of
the hierarchical model, and therefore it is much more flexible for
current and future elaborations of the basic model. Such
extensions include more general stick-breaking random measures,
and modelling dependence of the data on covariates, as for example
in \cite{duns}. A second advantage is that in principle  it allows
inference for the latent random measure $P$.

The implementation of the conditional method poses interesting
methodological challenges, since it requires the imputation of the
infinite-dimensional vectors $p$ and $Z$. The { modus operandi}
advocated in \cite{ishw:zarep:2000} is to approximate the vectors,
and thus the corresponding Dirichlet process prior, using some
kind of truncation, such as $p^{(N)}=(p_1,\ldots,p_N)$ and
$Z^{(N)}=(Z_1,\ldots,Z_N)$, where $N$ determines the degree of
approximation.  Although in some cases it is possible to control
the error produced by such truncations
\citep{ishw:zarep:2000,ishw:zarep:2002,ishw:jame:2001} it would be
desirable to avoid approximations altogether.

\def\MCMC{Markov chain Monte Carlo }
This paper introduces two implementations of the conditional
method which avoid any approximation. The proposed \MCMC
algorithms are very easily implemented
and can readily be
extended to more general stick-breaking models. The
implementation of the conditional method is achieved by{
retrospective sampling}, which is introduced in this paper. This
is a novel technique which facilitates exact simulation in finite
time in problems which involve infinite-dimensional processes. We
also show  how to use retrospective sampling in conjunction with
our \MCMC algorithms in order to sample from the posterior
distribution of functionals of $P$.

We identify a computational problem with the
conditional method. As a result of the weak identifiability imposed on the
labels of the mixture components, the posterior distribution of
the random measure $(p,Z)$ is multimodal. Therefore, the Gibbs
sampler has to visit all these different modes. This problem is
not eliminated with large datasets, since although the secondary
modes become smaller, the energy gap between the modes becomes
bigger, and thus the sampler can get trapped in low-probability areas
of the state-space of $(p,Z)$. We design tailored label-switching
moves which improve significantly the performance of our \MCMC
algorithm.

We also contrast our retrospective \MCMC algorithm
with state-of-the-art implementations of the marginal method for
non-conjugate models in terms of their Monte Carlo efficiency. In
particular we consider  the no-gaps algorithm of
\cite{mace:mull:1998} and  Algorithms 7 and 8 of
\cite{neal:2000}. This comparison sheds light on the relative
merits of the marginal and conditional method in models where they
can both be applied. We find that the marginal methods slightly
outperform the conditional.

Since the original submission of this
paper, the retrospective sampling ideas we introduce here have
been found very useful in extensions of the Dirichlet process
hierarchical model \citep{duns,grif:2006}, and
in the exact simulation of diffusion processes; see for example
\cite{besk:papa:robe:fear:2005}.

\section{Retrospective sampling from the Dirichlet process prior}
\label{sim-prior}

Consider the task of simulating a sample $X=(X_1,\ldots,X_n)$ from
the Dirichlet process prior (\ref{series}). Such a sample has the
property that $X_i \mid P \sim P$, $X_i$ is independent of $X_l$
conditionally on $P$ for each $l\neq i$, $i,l=1,\ldots,n$, and $P$
is the Dirichlet process with concentration parameter $\alpha$ and
base measure $H_\Theta$. This section introduces in a simple
context the technique of retrospective sampling and the two
different computational approaches, marginal and conditional, to
inference for Dirichlet processes. There are essentially two
different ways of obtaining $X$.

The sample
$X$ can be simulated directly from its marginal distribution. When
$P$ is marginalised the joint distribution of the $X_i$s is known
and it is  described by a P{\'o}lya urn scheme
\citep{blac:macq:1973,ferg:1973}. In particular, let $K_i$ be an
allocation variable associated with $X_i$, where $K_i=j$ if and
only if $X_i=Z_j$. Therefore, the $K_i$'s decide which component
of the infinite series (\ref{series}) $X_i$ is associated with.
Conditionally on $P$ the $K_i$'s are independent, with
\begin{equation}
\pr(K_i=j \mid p)=p_j,~~\mbox{for
all}~i=1,\ldots,n,~j=1,2,\ldots\,. \label{prior-k}
\end{equation}
The marginal prior of $K=(K_1,\ldots,K_n)$ is obtained by the
following P{\'o}lya urn scheme:
\begin{equation}
\label{polya}
\begin{split}
&\pr (K_i = j \mid K_1,\ldots,K_{i-1} )  =
 \frac{n_{i,j}}{(\alpha+i-1)}\,,  ~\textrm{if $K_l=j$ for some $l < i$}\,,  \\
& \pr (K_i \neq K_l ~\textrm{for all}~l < i \mid
K_1,\ldots,K_{i-1} )  =   \frac{\alpha}{(\alpha+i-1)} \, ,
\end{split}
\end{equation}
where $n_{i,j}$ denotes the size of the set $\{l < i: K_l=j\}$.
Thus, the probability that the $i$th sample is associated with the
$j$th component is proportional to the number of samples already
associated with $j$, whereas with probability
$\alpha/(\alpha+i-1)$ the $i$th sampled value is associated with a
new component. Note that, whereas in (\ref{prior-k}) the labels
$j$ are identifiable, in (\ref{polya}) the labels of the clusters
are totally arbitrary. Simulation of $X$ from its marginal
distribution  proceeds in following way. Set $K_1=1$, although any
other label could be chosen; simulate $\phi_1 \sim H_\Theta$; and set
$X_1=\phi_1$.  For $i>1$, let $c$ denote the number of existing
clusters; simulate $K_i$ conditionally on $K_1,\ldots,K_{i-1}$
according to the probabilities (\ref{polya}), where
$j=1,\ldots,c$; if $K_i=j$ then set $X_i=\phi_j$, and otherwise set
$c=c+1$, simulate $\phi_{c} \sim H_\Theta$ independently of any
previously drawn values, and set $X_i=\phi_{c}$. At the end of
this procedure $\phi_F=(\phi_1,\ldots,\phi_c)$ are the parameters
of the mixture components which are associated with at least one
data point. The indexing $1,\ldots,k$ is arbitrary and it is not
feasible to map these $\phi_j$'s to the $Z_j$'s in the definition
of the Dirichlet process in (\ref{series}).

Alternatively, $X$ can be simulated following a two-step
hierarchical procedure. Initially, a realisation of $P$ is
simulated, and then we simulate independently $n$ allocation
variables according to (\ref{prior-k}) and we set $X_i=Z_{K_i}$.
However, simulation of $P$ entails the generation of the infinite
vectors $p$ and $Z$, which is infeasible. Nevertheless, this
problem can be avoided with the following retrospective
simulation. The standard method for simulating from the discrete
distribution defined in (\ref{prior-k}) is first to simulate $U_i$
from a uniform distribution on $(0,1)$, and then to set $K_i=j$ if
and only if
\begin{equation}
\sum_{l=0}^{j-1}p_l < U_i \leq \sum_{l=1}^{j}p_l \ ,
\label{cond}
\end{equation}
where we define $p_0=0$; this is the inverse cumulative distribution
function method for discrete random variables; see for example
\nocite{ripl:1987}
Ripley (1987, \S 3.3).
 Retrospective sampling simply exchanges the order of simulation
between  $U_i$ and the pairs $(p_j,Z_j)$.  Rather than simulating
first $(p,Z)$ and then $U_i$ in order to check (\ref{cond}), we
first simulate the decision variable $U_i$ and then pairs
$(p_j,Z_j)$.  If for a given $U_i$ we need more $p_j$'s than we
currently have in order to check (\ref{cond}), then we go back and
simulate pairs $(p_j,Z_j)$ {
 `retrospectively'}, until (\ref{cond}) is satisfied.  The algorithm proceeds
 as follows.
\vspace{2mm}

\noindent
ALGORITHM 1. {\em Retrospective sampling from the Dirichlet process prior
\newline { \noindent Step {\rm 1}. Simulate $p_1$ and $Z_1$,
and set $N^{\ast}=1,~i=1$ and $p_0=0$.
\newline
%
Step {\rm 2}. Repeat the following until $i>n$.
 \vspace{0mm} \newline Step {\rm 2.1}. Simulate
$U_i \sim \Un[0,1]$ \newline Step {\rm 2.2}. If (\ref{cond}) is true for
some $k \leq N^{\ast}$ then
 set $K_i=k,~X_i=Z_k$, for $i=i+1$, and go to Step {\rm 2}.
\newline
Step {\rm 2.3}. If (\ref{cond}) is not true for any $k \leq N^{\ast}$
then set $N^{\ast}=N^{\ast}+1,~j=N^{\ast}$, simulate $p_j$ and
$Z_j$, and go to Step {\rm 2.2}.
 \newline
}\vspace{0mm}
}

\noindent In this notation, $N^{\ast}$ keeps track of how far into
the infinite sequence $\{(p_j,Z_j),~j=1,2,\ldots \}$ we have
visited during the simulation of the $X_i$'s. Note that the
retrospective sampling can be easily implemented because of the
Markovian structure of the $p_j$'s and the independence of the
$Z_j$'s. A similar scheme was advocated in
\cite{doss:1994}.

The previous simulation scheme illustrates the main principle
behind retrospective sampling: although it is impossible to
simulate an infinite-dimensional random object we might still be
able to take decisions  which depend on such objects  exactly
avoiding any approximations. The success of the approach will
depend on whether or not the decision problem can be formulated in
a way that involves finite-dimensional summaries, possibly of
random dimension, of the random object. In the previous toy
example we formulated the problem of simulating draws from the
Dirichlet process prior as the problem of comparing a uniform
random variable with partial sums of the Dirichlet random measure.
This facilitated the simulation of $X$ in finite time avoiding
approximation errors. In general, the retrospective simulation
scheme will require at the first stage simulation of both the
decision variable and certain finite-dimensional summaries of the
infinite-dimensional random object. Thus, at the second stage we
will need to simulate {retrospectively} from the distribution of
the random object conditionally on these summaries. This
conditional simulation will typically be much more elaborate than
the illustration we gave here.

We shall see in \Section \ref{cond-mcmc} that these ideas
extend to posterior simulation in a computationally feasible way.
However, the details of the method for posterior simulation
are far more complicated than the simple method given
above.

\section{The retrospective conditional method for posterior simulation}
\label{cond-mcmc}

\subsection{Posterior inference}
\label{posteriorinference}

When we fit (\ref{hier-cond})
to data $Y=(Y_1,\ldots,Y_n)$, there are a number of quantities about
which we may want to make posterior inference. These include
the classification variables $K_i$,  which can be used to classify
the data into clusters, the number of clusters in the population,
the  cluster parameters $\{Z_j: K_i=j ~\mbox{for at least
one}~i\}$ and the cluster probabilities $\{p_j: K_i=j ~\mbox{for
at least one}~i\}$, the hyperparameters $\alpha,\Theta$ and
$\lambda$, the predictive distribution of future data, and the
random measure $P$ itself. None of the existing methods can
provide samples from the posterior distribution of $P$ without
resorting to some kind of approximation; see for example
\cite{gelf:kott:2002} for some suggestions. However, exact
posterior simulation of finite-dimensional distributions and
functionals of $P$ might be feasible; see \Section
\ref{sec:non-lin}.

Markov chain Monte Carlo techniques have been developed for
sampling-based exploration of the posterior distributions outlined
above. The conditional method
\citep{ishw:zarep:2000} \nocite{ishw:jame:2001}
is based on an augmentation scheme in which the random
probabilities $p=(p_1,p_2,\ldots)$ and the component parameters
$Z=(Z_1,Z_2,\ldots)$ are imputed and the Gibbs sampler is used to
sample from the joint posterior distribution of $(K,p,Z)$
according to the full-conditional distributions of the variables.
However, the fact that $p$ and $Z$ have countably infinite
elements poses a major challenge. \cite{ishw:zarep:2000} advocated
an approximation of these vectors, and therefore an approximation
of the corresponding Dirichlet process prior. For instance  a
truncation of the Sethuraman representation (\ref{series}) could
be adopted, e.g.\@ $p^{(N)}=(p_1,\ldots,p_N)$ and
$Z^{(N)}=(Z_1,\ldots,Z_N)$, where $N$ determines the degree of
approximation.

In this section we show how to avoid any such approximation. The
first step of our solution is to parameterise in terms of
$(K,V,Z)$, where $V=(V_1,V_2,\ldots)$ are the beta random
variables used in the stick-breaking construction of $p$. We will
construct algorithms which effectively can return samples from the
posterior distribution of these vectors, by simulating iteratively
from the full conditional posterior distributions of $K$, $V$ and
$Z$. Note that we can replace direct simulation from the
conditional distributions with any updating mechanism, such as a
Metropolis-Hastings step, which is invariant with respect to these
conditional distribution. The stationary distribution of this more
general componentwise-updating sampler is still the joint
posterior distribution of $(K,V,Z)$.

We now introduce some notation.
For a given
configuration of the classification variables
$K=(K_1,\ldots,K_n)$, we
 define
\begin{equation*}
m_j = \sum_{i=1}^n 1_{\{K_i=j\}},~~j=1,2,\ldots,
\end{equation*}
to be the number of data points allocated to the $j$th component
of the mixture.
Moreover,  again for a given configuration of $K$ let
\begin{eqnarray}
I & = & \{1,2,\ldots\} \nonumber \\
I^{(\hbox{al})} & = & \{j \in I:m_j>0\}
\nonumber\\
I^{(\hbox{d})} & = & \{j \in I:m_j=0\}=I-I^{(
\hbox{al}
)}. \nonumber
\end{eqnarray}
Therefore, $I$  represents all components in the infinite mixture,
$I^{\al}$ is the set of all {`alive'} components and $I^{(\d)}$
the set of {`dead'} components, where we call a component `alive'
if some data have been allocated to it. The corresponding
partition of $Z$ and $V$ will be denoted by
$Z^{(\al)},Z^{(\d)},V^{(\al)}$ and $V^{(\d)}$.

\comment{The second step of our approach is to note that to
perform most statistical analyses of interest it suffices to have
samples from the posterior distribution of $K$, $Z^{(\al)}$ and
$V^{(\al)}$. Indeed we will show that given such samples we can
estimate most quantities of interest. Therefore, our final aim
will be to design a three-component Gibbs sampler which simulates
iteratively $K$, $Z^{(\al)}$ and $V^{(\al)}$ from their full
conditional distributions. In principle we can replace direct
simulation from the conditional distributions with any updating
mechanism, such as a Metropolis-Hastings step, which is invariant
with respect to the corresponding conditional distribution. We
will refer to this more general algorithm as a
{componentwise-updating sampler}. }
 The implementation of our \MCMC
algorithms relies on the results contained in the following
Proposition, which describes the full conditional posterior
distributions of $Z,V$ and $K$.
%

\medskip
\noindent
{\bf Proposition 1.} \ \
{\em
Let $\pi(z \mid \Theta)$ be the density of
$H_\Theta({d}z)$.  Conditionally on $(Y,K,\Theta,\lambda)$,
$Z$ is independent of $(V,\alpha)$ and it consists of
conditionally independent elements with
\begin{equation*}
Z_j \mid  Y,K,\Theta,\lambda \sim  \left \{ \begin{array}{lll}
 H_{\Theta},~\mbox{for all}~j \in I^{(\d)}  \\ \\
\prod_{i:K_i=j} f(Y_i \mid Z_j,\lambda) \pi(Z_j \mid
\Theta)~\mbox{for all}~j \in I^{(\al)}\,,
 \end{array}
\right. \label{zjs}
\end{equation*}
 Conditionally on $(K,\alpha)$, $V$ is independent of
 $(Y,Z,\Theta,\lambda)$ and it consists of conditionally
independent elements with
\begin{equation*}
V_j \mid K,\alpha  \sim {\mathrm{Be}}\left
(m_j+1,n-\sum_{l=1}^{j}m_l+\alpha \right) ~\mbox{for
all}~j=1,2,\ldots \,. \label{vjs}
\end{equation*}
Conditionally on $(Y,V,Z,\lambda)$, $K$ is independent of
$(\Theta,\alpha)$ and it  consist of conditionally independent
elements with
\begin{equation*}
\mathrm{pr}\{K_i=j \mid Y,V,Z,\lambda\}  \propto  p_j f(Y_i \mid
Z_j,\lambda),~j=1,2,\ldots \,. \label{ki-post}
\end{equation*}
%
%
}

\noindent
The proof of the Proposition follows directly from the conditional
independence structure in the model and the stick-breaking
representation of the Dirichlet process;  see
\cite{ishw:jame:2003} for a proof in a more general context.

The conditional independence structure in the model effects
the simultaneous sampling of any finite subset of pairs
$(V_j,Z_j)$ according to their full conditional distributions.
Simulation of $(Z_j,V_j)$ when $j \in I^{(\d)}$ is trivial. When
dealing with non-conjugate models, the distribution of $Z_j$ for
$j \in I^{(\al)}$ will typically not belong to a known family and a
Metropolis-Hastings step might be used instead to carry out this
simulation. Note that samples from the full conditional
posterior distribution of $p_j$'s are obtained using samples of
$(V_h,h \leq j)$ and the stick-breaking representation in
(\ref{hier-cond}).
\comment{
 If $j \in
I^{(\d)}$ then simply simulate $Z_j \sim H_{\Theta}$. When $\pi(z
\mid \Theta)$ and $f(y \mid z,\lambda)$ form a conjugate pair for
$z$, then the conditional distribution of $Z_j$ is of known form
for all $j \in I^{(\al)}$, and it can be sampled easily. For
non-conjugate models rejection sampling or a Metropolis-Hastings
step can be used to update $Z_j, j \in I^{(\al)}$.  Simulation of
the $p_j$'s conditionally on $Y$ and $K$ can be easily achieved
using the stick-breaking representation (\ref{ps-seq}) and
Proposition 1
above:  first simulate
$V_1,\ldots,V_j$  from (\ref{vjs}), and then apply (\ref{ps}). If
another draw $p_h$ is required then, if $h<j$, apply (\ref{ps})
using the simulated $V_l,l=1,\ldots,h$, whereas if $h>j$, simulate
$V_{j+1},\ldots,V_h$ according to (\ref{vjs}) and apply
(\ref{ps}). }

The $K_i$'s are conditionally independent given $Z$ and $V$.
However, the normalising constant of the full conditional
probability mass function of each $K_i$ is intractable:
\begin{equation*}
c_i=\sum_{j=1}^{\infty} p_j  f(Y_i \mid Z_j,\lambda)\,.
\label{norm}
\end{equation*}
The intractability stems from the fact that an infinite sum of
random terms needs to be computed. At this stage one could resort
to a finite approximation of the sums, but we wish to avoid
this.  The unavailability of the normalising constants renders
simulation of the $K_i$'s highly nontrivial. Therefore, in order
to construct a conditional \MCMC algorithm we need to find ways of
sampling from the conditional distribution of the $K_i$'s.

\comment{Therefore, the retrospective simulation scheme we
introduced in \Section \ref{sim-prior} for simulation from the
Dirichlet process prior cannot be applied here. An alternative
retrospective scheme is required, and \Section \ref{sec:quasi} and
\Section \ref{sec:retro-unnorm} introduce two algorithms which can
be used to achieve sampling of $K$ according to its full
conditional distribution.}

\subsection{A retrospective quasi-independence Metropolis-Hastings sampler
for the allocation  variables} \label{sec:quasi}

One simple and effective way of avoiding the  computation of the
normalising constants $c_i$ is to replace direct simulation of the
$K_i$s with a Metropolis-Hastings step. Let $k=(k_1,\ldots,k_n)$
denote a configuration of $K=(K_1,\ldots,K_n)$, and let
\begin{equation*}
\kmax = \max_i k_i \label{kmax}
\end{equation*}
be the maximal element of the vector $k$. We assume that we have
already obtained samples from the conditional posterior
distribution of $\{(V_j,Z_j): j\leq \kmax\}$ given $K=k$. Note
that the distribution of $(V_j,Z_j)$ conditionally on $Y$ and
$K=k$ is simply the prior, $V_j \sim \Be(\alpha,1), Z_j \sim
H_\Theta$, for any $j>\kmax$.

We will describe an updating scheme $k \mapsto k^*$ which is
invariant with respect to the full conditional distribution of $K
\mid Y,Z,V,\lambda$. The scheme is a composition of $n$
Metropolis-Hastings steps which update each of the $K_i$s in turn.
Let
$$k(i,j)=(k_1,\ldots,k_{i-1},j,k_{i+1},\ldots,k_n)$$
be the vector produced from $k$ by substituting the $i$th element
by $j$. When updating $K_i$, the sampler proposes to move from $k$
to $k(i,j)$, where the proposed $j$ is generated  from the
probability mass function
\begin{equation}
q_i(k,j) \propto \left \{ \begin{array}{ll} p_j f(Y_i \mid
Z_j,\lambda), &
~~~\textrm{for}~j\leq \kmax  \\ \\
M_i(k) p_j, & ~~~\textrm{for}~j> \kmax\,.
 \end{array}
\right. \label{mcmc-prop}
\end{equation}
The  normalising constant of (\ref{mcmc-prop}) is
\begin{equation*}
\qcon(k) = \sum_{j=1}^{\kkmax}p_j f(Y_i \mid
Z_j,\lambda)+M_i(k)\left (1-\sum_{j=1}^{\kkmax}p_j\right),
\label{norm-mcmc}
\end{equation*}
which can be easily computed given $\{(p_j,Z_j):j \leq \kmax\}$.
Note that, for $j \leq \kmax$, $q_i(k,j) \propto \pr(K_i=j \mid
Y,V,Z,\lambda)$, while, for $j>\kmax$, $q_i(k,j) \propto \pr
(K_i=j \mid V)$. Here $M_i(k)$  is a user-specified parameter which
controls the probability of proposing $j
> \kmax$, and its choice will be discussed in \Section
\ref{sec:acc}.

According to this proposal distribution the $i$th data point is
proposed to be re-allocated to one of the  alive clusters $j \in
I^{(\al)}$ with probability proportional to the conditional
posterior probability $\pr(K_i=j \mid Y,V,Z,\lambda)$. Allocation
of the $i$th data point to a new component can be accomplished in
two ways:  by proposing $j \in I^{(\d)}$ for $j \leq \kmax$
according to the conditional posterior probability mass function
and by proposing $j \in I^{(\d)}$ for $j> \kmax$ according to the
prior probability mass function. Therefore, a careful calculation
yields that the Metropolis-Hastings acceptance probability of the
transition from $K=k$ to $K=k(i,j)$ is
%
%
\begin{equation*}
\label{eq:accept-prob} \alpha_i\{k,k(i,j)\} = \left \{
\begin{array}{lllll}
1, &\hspace{-3mm} \small \textrm{if}~j\leq \kmax~\mbox{and}~\kkkmaxp= \kmax \\ \\
\min \left \{1,
{
{
\qcon(k)M\{k(i,j)\}
\over
\qcon\{k(i,j)\}f(Y_i \mid Z_{k_i},\lambda)
}
}
\right \}, &\hspace{-3mm} \small
\textrm{if}~j \leq
  \kmax~\mbox{and}~\kkkmaxp< \kmax \\ \\
\min \left \{1,
{
{
\qcon(k)f(Y_i \mid Z_{j},\lambda)
\over
\qcon\{k(i,j)\}M(k)
}
}
\right \}, &\hspace{-3mm}
\small \textrm{if}~j
>\kmax.
 \end{array}
\right.
\end{equation*}
%
{Note that} proposed  re-allocations to a component $j \leq \kmax$
are accepted with probability 1 provided that $\kmaxp=\kmax$. If
the proposed move is accepted, we set $k=k(i,j)$ and proceed to
the updating of $K_{i+1}$. The composition of all these steps
yields the updating mechanism for $K$.

Simulation from the proposal distribution is achieved by
retrospective sampling. For each $i=1,\ldots,n$, we simulate $U_i
\sim \Un[0,1]$ and propose to set $K_i=j$, where $j$ satisfies
\begin{equation}
\sum_{l=0}^{j-1} q_i(k,l) \ < U_i  \ \leq \sum_{l=1}^{j} q_i(k,l),
\label{cond-post}
\end{equation}
with $ q_i(k,0) \eqdef 0$. If (\ref{cond-post}) is not satisfied
for any $j \leq \kmax$, then we start checking the condition for
$j>\kmax$ until it is satisfied. This will require the values of
$(V_{l},Z_{l}),l
> \kmax$. If these values have not already been simulated in the previous
steps of the algorithm, they are simulated retrospectively  from
their prior distribution when they become needed.

We therefore have a retrospective \MCMC algorithm for sampling
from the joint posterior distribution of $(K,V,Z)$, which is
summarised below.  
\newline

\noindent
ALGORITHM 2.
{\em Retrospective Markov chain Monte Carlo
\newline Give an
initial allocation $k=(k_1,\ldots,k_n)$, and  set $N^*=\kmax$
\newline
Step {\rm 1}. Simulate $Z_j$ from its conditional posterior, $j \leq
\kmax$.
\newline Step {\rm 2.1} Simulate $V_j$ from its conditional
posterior, $j \leq \kmax$. \newline $~$Step {\rm 2.2} Calculate
$p_j=(1-V_1)\cdots(1-V_{j-1})V_j,~j \leq \kmax$. \newline Step
{\rm 3.1}. Repeat the following until $i>n$. \vspace{0mm}
\newline Step {\rm 3.2}. Simulate $U_i \sim \Un[0,1]$ \newline Step
{\rm 3.3.1}
If (\ref{cond-post}) is true for some $j \leq N^{\ast}$ then
set $K_i=j$
with probability $\alpha_i\{k,k(i,j)\}$, otherwise leave it
unchanged. Set $i=i+1$ and
go to Step {\rm 3.1}. \newline
Step {\rm 3.3.2} If (\ref{cond-post})  is  not true for any $j \leq N^{\ast}$,  
set $ N^{\ast}=N^{\ast}+1,~j=N^{\ast}$. Simulate $(V_j,Z_j)$ from
the prior, set $p_j=(1-V_1)\cdots (1-V_{j-1})V_j$ and go to Step
{\rm 3.3.1} \newline  Step {\rm 3.4}. Set $N^*=\max\{k\}$ and go to Step {\rm
1}.
\vspace{2mm}}
\label{pg:retro-mcmc}
%

\noindent Note that both $\kmax$ and $N^*$ change during the
updating of the $K_i$'s, with $N^\ast \geq \kmax$.  Thus, the
proposal distribution  is adapting itself to improve approximation
of the target distribution. At the early stages of the algorithm
$N^*$ will be large, but as the cluster structure starts being
identified by the data then $N^*$ will take much smaller values.
Nevertheless, the adaptation of the algorithm does not violate the
Markov property. It is recommended to update the $K_i$'s in a
random order, to avoid using systematically less efficient
proposals for some of the variables.

The algorithm we have constructed updates all the allocation
variables $K$ but only a random subset of $(V,Z)$, to be precise
$\{(V_j,Z_j),j\leq N^*\}$. However, pairs of the complementary set
$\{(V_j,Z_j),j> N^*\}$ can be simulated when and if they are
needed retrospectively from the prior distribution. This
simulation can be performed off-line after the completion of the
algorithm. In this sense, our algorithm is capable of exploring
the joint distribution of $(K,V,Z)$.

\subsection{Accelerations of the main algorithm}
\label{sec:acc}

 There are several simple modifications of the
retrospective \MCMC algorithm which can improve significantly its
Monte Carlo efficiency.

We first discuss the choice of the user-specified parameter
$M_i(k)$ in (\ref{mcmc-prop}). This parameter relates to the
probability, $\rho$ say, of proposing   $j
> \kmax$, where
\begin{equation*}
\rho=\frac{\left(1-\sum_{j=1}^{\kkmax}p_j\right)M_i(k)}{c_i(\kmax)}\,.
\label{prop-prob}
\end{equation*}
If it is desired to propose components $j>\kmax$ a specific
proportion of the time then the equation above can be solved for
$M_i(k)$. For example, $\rho=\alpha/(n-1)$ is the probability of
proposing new clusters in Algorithm 7 of \cite{neal:2000}. We
recommend a choice of $M_i(k)$ which guarantees that the
probability of proposing $j>\kmax$ is greater than the prior
probability assigned to the set $\{j:j>\kmax\}$. Thus $M_i(k)$
should satisfy
\begin{equation*}
\sum_{j=1}^{\kkmax}p_j f(Y_i \mid
Z_j,\phi)+M_i(k)\left(1-\sum_{j=1}^{\kkmax}p_j\right) \leq M_i(k).
\label{m-ineq}
\end{equation*}
This inequality is satisfied by setting
\begin{equation}
M_i(k)=\max \left \{f(Y_i \mid Z_j,\phi),j\leq \kmax \right\}.
\label{mi}
\end{equation}
Since (\ref{mcmc-prop}) resembles an independence sampler for
$K_i$, (\ref{mi}) is  advisable from a theoretical perspective.
\cite{meng:twee:1996} have shown that an independence sampler is
geometrically ergodic if and only if the tails of the proposal
distribution are heavier than the tails of the target
distribution. The choice of $M_i(k)$ according to (\ref{mi})
ensures that the tails of the proposal $q_i(k,j)$ are heavier than
the tails of the target probability $\pr\{K_i=j \mid
Y,p,Z,\lambda\}$. Note that when $M_i(k)$ is chosen according to
(\ref{mi}) then $\rho$ is random. In simulation studies we have
discovered that the distribution of $\rho$ is very skewed and the
choice according to (\ref{mi}) leads to a faster mixing algorithm
than alternative schemes with fixed $\rho$.

Another interesting possibility is to update  $Z$ and $V$ after
each update of the allocation variables.  Before Step 3.2 of
Algorithm 2  we simulate $Z^{(\d)}$ from the prior and leave
$Z^{(\al)}$ unchanged.  Moreover, we can synchronise $N^\ast$ and
$\kmax$. Theoretically, this is achieved by pretending to update
$\{V_j,j
> \kmax\}$ from the prior before Step 3.2. In practice, the only
adjustment to the existing algorithm is to set $N^\ast=\kmax$
before Step 3.2.  These extra updates have the computational
advantage of storing only $Z^{(\al)}$ and $\{V_j, j \leq \kmax \}$.
In addition, simulations have verified that these extra updates
improve significantly the mixing of the algorithm. Morover, one
can allow more components $j \in I^{(\d)}$ to be proposed according
to the posterior probability mass function by changing $\kmax$ in
(\ref{mcmc-prop}) to $\kmax+l$, for some fixed integer $l$. In
that case the acceptance probability (\ref{eq:accept-prob}) needs
to be slightly modified, but we have not found gains from such
adjustment.

\subsection{Multimodality and label-switching moves}

The most significant and crucial modification of the algorithm we
have introduced is the addition of label-switching moves. These
moves will have to be included in any implementation of the
conditional method which updates the allocation variables one at a
time. The augmentation of $p$ in the conditional method makes the
components in the infinite mixture (\ref{hier-cond}) weakly
identifiable, in the sense that $E\{p_j \} \geq E\{p_l \}$ for any
$j \geq l$, but there is nonnegligible prior probability that
$p_l>p_j$, in particular when $|l-j|$ is small. As a result the
posterior distribution of $(p,Z)$ exhibits multiple modes. In
order to highlight this phenomenon we consider below a simplified
scenario, but we refer to \Section \ref{comp}
for an illustration in the context of posterior inference for a specific
non-conjugate Dirichlet process hierarchical model.

\begin{figure}[h]
\begin{center}
\begin{tabular}{lll}
(a) & (b) & (c) \\
\psfig{figure=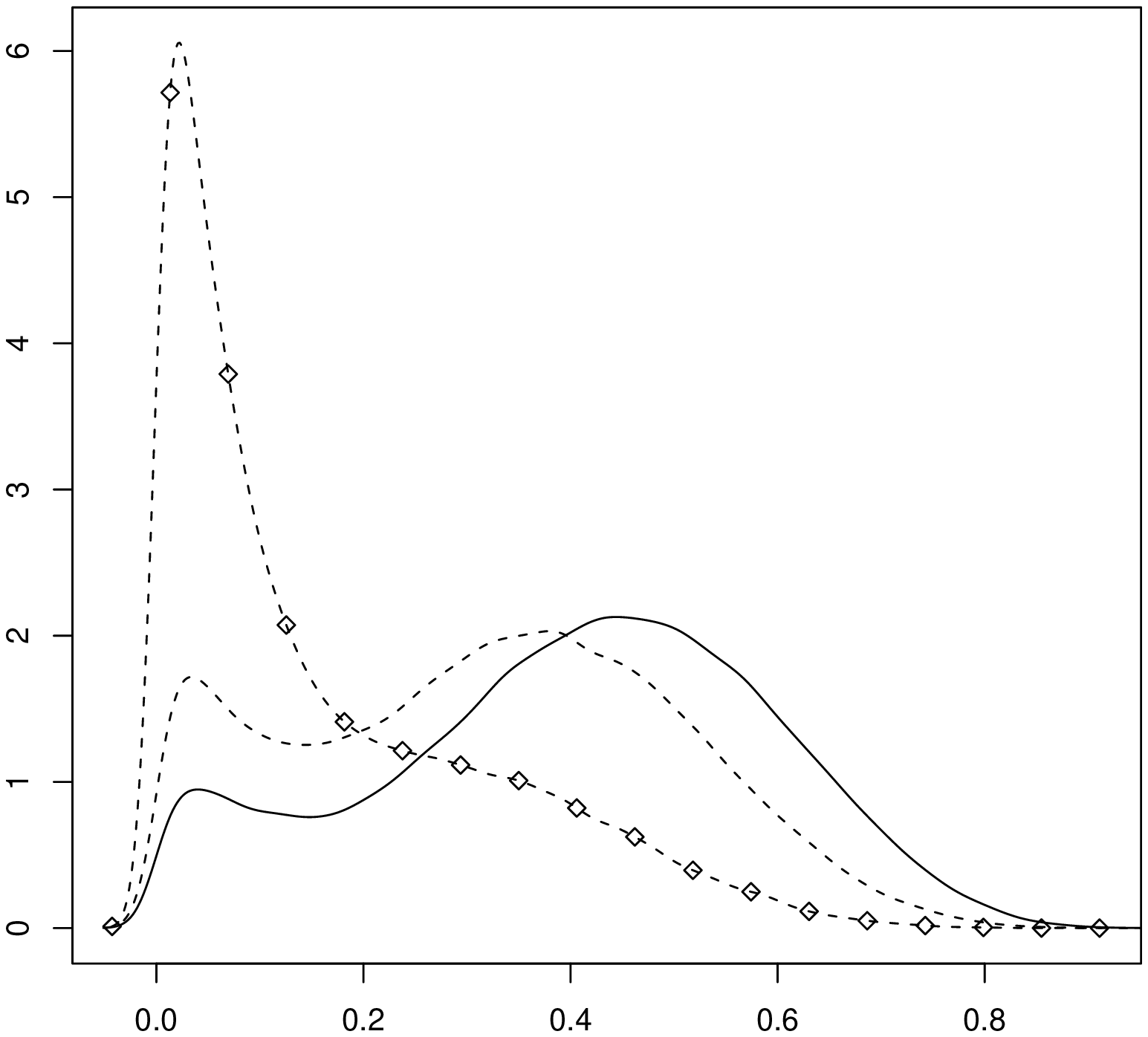,height=2in,width=2.5in,angle=270} &
\psfig{figure=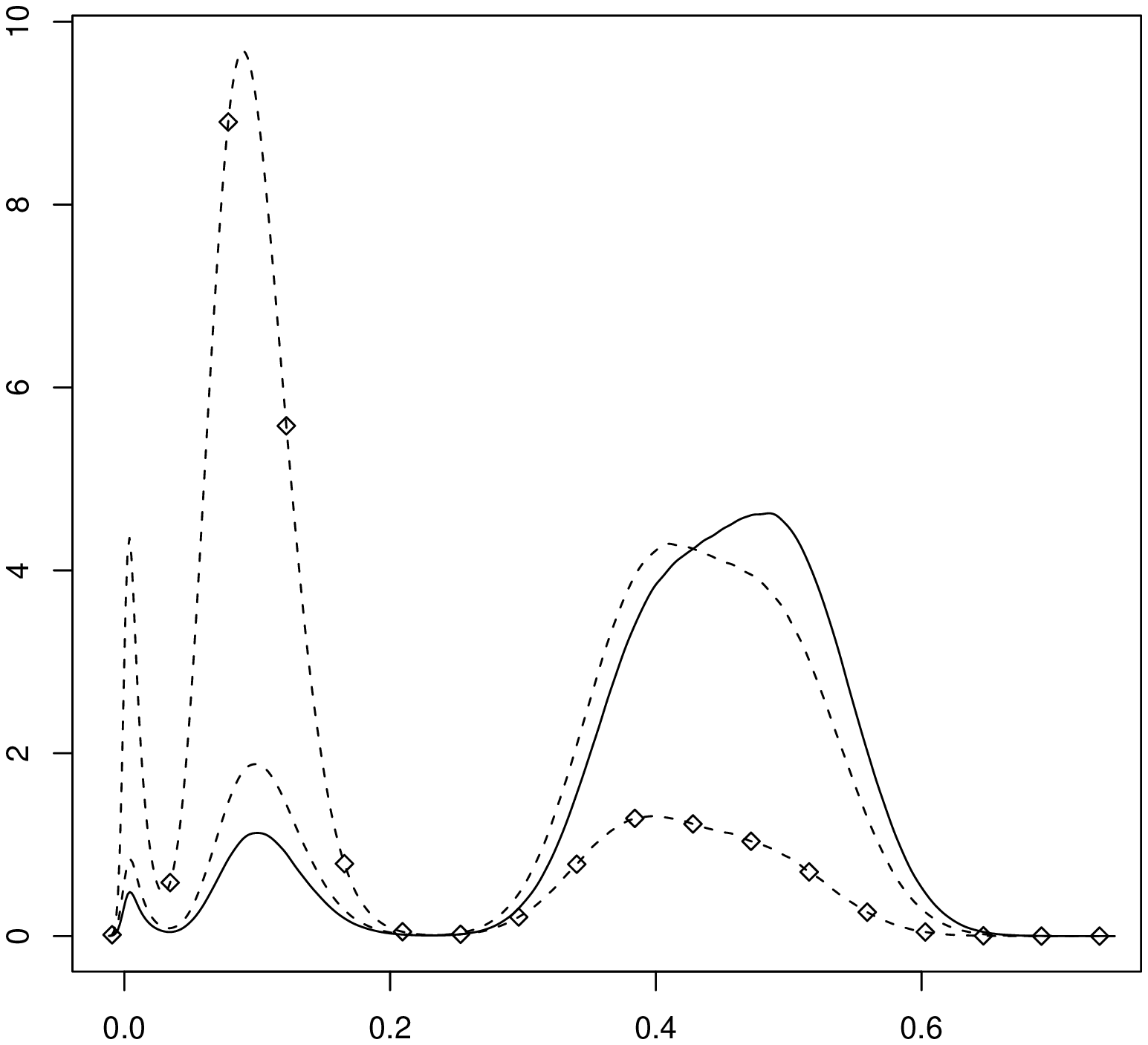,height=2in,width=2.5in,angle=270} &
\psfig{figure=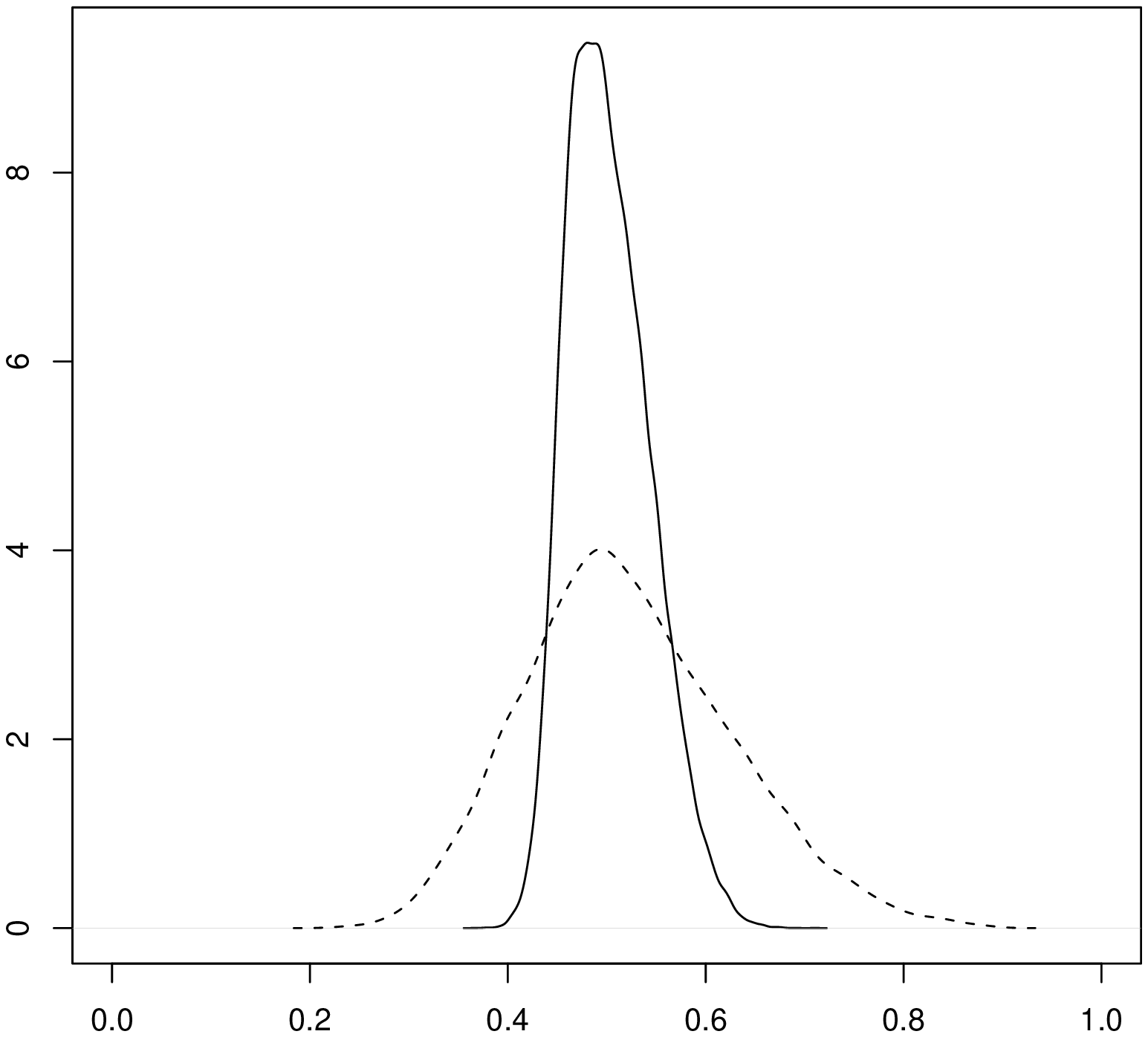,height=2in,width=2.5in,angle=270}
\end{tabular}
\end{center}
\caption{Posterior densities of $p_1$ (solid),
$p_2$ (dashed) and $p_3$ (dashed with diamonds),
corresponding to (a) a dataset with $n=10$ separated into three
clusters of sizes $n_1=5,n_2=4,n_1$ and (b)
a dataset with $n=100,n_1=50,n_2=40$ and $n_3=10$. (c) shows
 the posterior
density of $\max_j\{p_j\}$ for $n=10$ (dashed) and $n=100$
(solid).} \label{fig:multi}
\end{figure}

In particular, assume that we have actually observed a sample of
size $n$ from the Dirichlet process $X=(X_1,\ldots,X_n)$ using
the notation of \Section \ref{sim-prior}. This is the limiting
case of (\ref{hier-cond}) where the observation density $f(y \mid
z,\lambda)$ is a point mass at $z$. In this case we directly
observe the allocation of the $X_i$'s into $c$, say, clusters, each
of size $n_l$, say, where $\sum_{l=1}^c n_l=n$. The common value
of the $X_i$'s in each cluster $l$ gives precisely the parameters
$\phi_l$ of the corresponding cluster $l=1,\ldots,c$. However,
there is still uncertainty regarding the component probabilities
$p_j$ of the Dirichlet process which generated the sample, and the
index of the component for which each cluster has been generated. Let
$K_l,l=1,\ldots,c$ denote these indices for each of the clusters.
We can construct a Gibbs sampler for exploring the posterior
distribution of $p$ and $(K_1,\ldots,K_c)$: this is a nontrivial
simulation, and a variation of the retrospective \MCMC algorithm has
to be used. Figure \ref{fig:multi} shows the posterior densities
of $(p_1,p_2,p_3)$ when $n=10$ and $c=3$ with $n_1=5,n_2=4,n_3=1$,
and when $n=100$ and $c=3$ with $n_1=50,n_2=40,n_3=10$. In both
cases we took $\alpha=1$.

Note that the posterior distributions of the $p_j$s exhibit
multiple modes, because the labels in the
mixture are only weakly identifiable. Note that the secondary
modes become less prominent for larger samples, but the energy gap
between the modes increases. In contrast, the probability of
the largest component, $\max\{p_j\}$, has a  unimodal posterior
density. This is shown in the right panel of Figure
\ref{fig:multi}(c); simulation from this density has been achieved by
retrospective sampling, see  \Section \ref{sec:non-lin}.

In general, in the conditional method the \MCMC algorithm has to
explore multimodal posterior distributions as those shown in
Figure \ref{fig:multi}. Therefore, we need to add label-switching
moves which assist the algorithm to jump across modes. This is
particularly important for large datasets, where the modes are
separated by areas of negligible probability. A careful inspection
of the problem suggests two types of move. The first proposes to
change the labels $j$ and $l$ of two randomly chosen components
$j,l \in I^{(\al)}$. The probability of such a change is accepted
with probability $\min\{1, (p_j/p_l)^{m_l-m_j} \}$. This proposal
has high acceptance probability if the two components have similar
weights; the probability is indeed 1 if $p_j=p_l$. On the other
hand, note that the probability of acceptance is small when
$|m_l-m_j|$ is close to 0. The second move proposes to change the
labels $j$ and $j+1$ of two neighbouring components but at the
same time to exchange $V_j$ with $V_{j+1}$. This change is
accepted with probability
$\min\{1,(1-V_{j+1})^{m_{j}}/(1-V_j)^{m_{j+1}} \}$. This proposal
is very effective for swapping the labels of very unequal
clusters. For example, it will always accept when $m_j=0$. On the
other hand, if $|p_j-p_{j+1}|$ is small, then the proposal will be
rejected with high probability. For example, if $V_j \bumpeq 1/2$,
$V_{j+1} \bumpeq 1$, and $m_j \bumpeq m_{j+1}$, then the proposal
attempts to allocate $m_{j}$ is allocated to a component of
negligible weight, $\prod_{l=1}^{j-1}(1-V_l) (1-V_{j+1})V_j$.
Thus, the two moves we have suggested are complementary to each
other.

Extensive simulation experimentation has revealed that such moves
are crucial. The integrated autocorrelation time, see \Section
\ref{comp} for definition, of the updated variables can be reduced
by as much as 50-90\%.

The problem of multimodality has also been addressed in a recent paper by
\cite{Porteous06}.

\subsection{Exact retrospective sampling of the allocation variables} \label{sec:retro-unnorm}

It is possible to avoid the Metropolis-Hastings step and simulate
directly from the conditional posterior distribution of the
allocation variables.  This is facilitated by a general
retrospective simulation scheme for sampling discrete random
variables whose probability mass function depends on the Dirichlet
random measure.

We can formulate the problem as one of
simulating a discrete random variable $J$, according to the
distribution specified by the probabilities
\begin{equation}
\label{eq:gen-retro} r_j:=\frac{p_j f_j}{\sum_{j=1}^\infty p_j
f_j}, ~j=1,2,\ldots\,.
\end{equation}
The $f_j$'s are assumed to be positive and independent random
variables, also independent of the $p_j$'s, which are given by the
stick-breaking rule in (\ref{hier-cond}), although more general
schemes can be incorporated in our method. The retrospective
sampling scheme of \Section \ref{sim-prior} cannot be applied
here, since the normalising constant of (\ref{eq:gen-retro}),
$c=\sum p_j f_j$, is unknown. In the specific application we have
in mind, $f_j=f(Y_i \mid Z_j,\lambda)$ for each allocation
variable $K_i$ we wish to update.

Nevertheless, a retrospective scheme can be applied if we can
construct two bounding sequences $\{c_l(k)\} \uparrow c$ and
$\{c_u(k)\} \downarrow c$, where $c_l(k)$ and $c_u(k)$ can be
calculated simply on the basis of $(p_1,Z_1), \ldots ,(p_k,Z_k)$.
Let $r_{u,j}(k) = r_j/c_l(k)$ and $r_{l,j}(k) = r_j/c_u(k)$. Then
we first simulate a uniform $U$ and then we set $J = j$ when
\begin{equation}
\label{eq:cond-giibs} \sum_{m=1}^{j-1}r_{u,m}(k) \le U \le
\sum_{m=1}^{j}r_{l,m}(k)\,.
\end{equation}
In this algorithm $k$ should be increased, and the $p_j$'s and $f_j$'s
simulated retrospectively, until (\ref{eq:cond-giibs}) can be
verified for some $j \leq k$.

Therefore, simulation from unnormalised discrete probability
distributions is feasible provided the bounding sequences can be
appropriately constructed. It turns out that this construction is
generally very challenging. In this paper we will only tackle the
simple case in which there exists a constant $M<\infty$ such that
$f_j<M$ for all $j$ almost surely. In our application this
corresponds to $f(y \mid z,\lambda)$ being bounded in $z$. In that
case one can simply take
\begin{eqnarray*}
c_l(k) & = & \sum_{j=1}^k p_j f_j \\
c_u(k) & = & c_l(k)+ M \left(1-\sum_{j=1}^k p_j \right)\,.
\end{eqnarray*}
When $\lim\sup f_j = \infty$ almost surely, an alternative
construction has to be devised, which involves an appropriate
coupling of the $f_j$'s. This construction is elaborate and
mathematically intricate, and
will be reported elsewhere.

\subsection{Posterior inference for functionals of Dirichlet processes}
\label{sec:non-lin}

 Suppose we are interested in estimating the
posterior distribution of
\begin{equation*}
\label{eq:non-lin-dp} \mathcal{I}=\int_{\mathcal{X}}g(x)P(dx),
\end{equation*}
for some real-valued function $g$, where $\mathcal{X}$ denotes the
space on which the component parameters $Z_j$ are defined and  $P$
is the Dirichlet random measure (\ref{series}), given data $Y$
from the hierarchical model (\ref{hier-cond}). Our implementation
of the conditional method shows that simulation from the posterior
distribution of $\mathcal{I}$ is as difficult as the
simulation from its prior. Let $\{(V_j,Z_j),j\leq \kmax\}$ be a
sample obtained using the retrospective \MCMC algorithm. We assume
that the sample is taken when the chain is `in stationarity',
i.e.\@ after a sufficient number of initial iterations. Given the
$V_j$'s we can compute the corresponding $p_j, j \leq \kmax$.
Recall that $Z_j \sim H_\Theta$ and $V_j \sim \Be(1,\alpha)$ for
all $j>\kmax$. Then we have the following representation for the
posterior distribution of $\mathcal{I}$: a draw from $\mathcal{I}
\mid Y$ can be represented as
$$ \sum_{j=1}^{\tiny \kmax} f(Z_j) p_j+ \sum_{j={\tiny \kmax}+1}^{\infty}
f(Z_j) p_j\, {=} \,\sum_{j=1}^{\tiny \kmax} f(Z_j) p_j +
\frac{I}{\prod_{l=1}^{\tiny \kmax}(1-V_l)} \,.
$$
This is equality in distribution, where $I$ is a draw from from
the prior distribution of $\mathcal{I}$.
Inference for linear functionals of the Dirichlet process was
initiated in \cite{cifa:rega:90} and simulation aspects have been
considered, for example in
\cite{gugl:twee:2001} and \cite{gugl:holm:walk:2002}.

Our algorithm can also be used for the simulation of
non-linear functionals under the posterior distribution.
As an illustrative example consider
posterior simulation of the predominant species corresponding to
$J$ such that $p_J \geq p_l$ for all $l=1,2,\ldots$. This can be
achieved as follows given a sample $\{(V_1,Z_1),j \leq \kmax\}$
obtained with any of our conditional \MCMC algorithms. Let
$J=\mbox{arg}\max_{\{j \leq N^\ast\}} p_j$, where $N^\ast=\kmax$.
Then, it can be seen that if $1-\sum_{j=1}^{N^\ast} p_j <p_J$ then
$J$ is the predominant specie. Therefore, if $1-\Pi_{N^\ast}>p_J$,
we repeat the following procedure until the condition is
satisfied: increase $N^\ast$, simulate pairs
$(Z_{N^\ast},V_{N^\ast})$ from the prior, and compute $J$. This
procedure was used to obtain the results in
Figure \ref{fig:multi}(c).


\subsection{Inference for hyperparameters}
\label{sec:hyper}

A simple modification of the retrospective algorithms described above
provides the computational machinery needed for Bayesian inference
for the hyperparameters $(\Theta,\alpha,\lambda)$. If we assume that
appropriate prior distributions have been chosen, the aim is to
simulate the hyperparameters according to their full conditional
distributions, thus adding one more step in the retrospective
\MCMC algorithm.

Sampling of $\lambda$ according to its full conditional
distribution is standard. At first glance, sampling of
$(\Theta,\alpha)$ poses a major complication. Note that, because of
the hierarchical structure, $(\Theta,\alpha)$ are independent of
$(Y,K)$ conditionally upon $(V,Z)$.  However, $(V,Z)$ contains an
infinite amount of information about $(\alpha,\Theta)$. Therefore,
an algorithm which updated successively $(V,Z,K)$ and
$(\Theta,\alpha)$ according to their full conditional
distributions would be reducible. This type of convergence problem
is not uncommon when \MCMC is used to infer about hierarchical
models with hidden stochastic processes; see
\cite{paprobsk,paprobsk2} for reviews.

However, the conditional independence structure in $Z$ and $V$
can be used to circumvent
the problem. In effect, instead of updating $(\Theta,\alpha)$
conditionally upon $(Z,V)$ we can jointly update
$(\Theta,\alpha,Z^{(\d)},V^{(\d)})$ conditionally upon
$(Z^{(\al)},V^{(\al)})$. In practice, we only need to simulate
$(\Theta,\alpha)$ conditionally on $(Z^{(\al)},V^{(\al)})$. That
poses no complication since $(Z^{(\al)},V^{(\al)})$  contains only
finite amount of information about the hyperparameters.  It is
worth mentioning that a similar technique for updating the
hyperparameters was recommended by \cite{mace:mull:1998} for the
implementation of their no-gaps algorithm.

\section{Comparison between marginal and conditional methods}
\label{comp}

In this section we attempt a comparison in terms of Monte Carlo
efficiency between the retrospective \MCMC algorithm
and state-of-the-art implementations of the marginal approach. To
this end we have carried out a large simulation study part of
which is presented later in this section. The methods are tested
on the non-conjugate model where $f(y \mid z)$ is a Gaussian
density with $z=(\mu,\sigma^2)$ and $H_\Theta$ is the product
measure $\N(\mu,\sigma^2_z)\times \IGa(\gamma,\beta)$; there are
no further parameters $\lambda$ indexing $f(y \mid z)$,
$\Theta=(\mu,\sigma_z^2,\gamma,\beta)$ and $\IGa(\gamma,\beta)$
denotes the inverse Gamma distribution with density proportional
to $x^{-(\gamma+1)}e^{-\beta x}$. Note that, in a
density-estimation context, this model allows for local smoothing; see for
example \cite{mull:erka:west:1996} and \cite{green:rich:2001}.

An excellent overview of different implementations of the marginal
approach can be found in \cite{neal:2000}. The most successful
implementations for non-conjugate models are the so-called no-gaps
algorithm of \cite{mace:mull:1998} and Algorithms 7 and 8 of
\cite{neal:2000}, which were introduced to mitigate against
certain computational inefficiencies of the no-gaps algorithm.

Received \MCMC { wisdom} suggests that marginal samplers ought to
be preferred to conditional ones. Some limited theory supports
this view; see in particular \cite{liu:1994}.
However, it is common for marginalisation to destroy conditional
independence structure which usually assists the conditional
sampler, since conditionally independent components are
effectively updated in one block. Thus, it is difficult {a priori}
to decide which approach is preferable.

In our comparison we have considered different simulated datasets
and different prior specifications. We have simulated 4 datasets
from two models. The `lepto 100' and the `lepto 1000' datasets
consist respectively of 100 and 1000 draws from the unimodal
leptokurtic mixture, $0.67 \N(0,1)+0.33 \N(0.3,0.25^2)$. The
`bimod 100' (`bimod 1000') dataset consists of 100 (1000) draws
from the bimodal mixture, $0.5\N(-1,0.5^2)+0.5\N(1,0.5^2)$; we
have chosen these datasets following \cite{green:rich:2001}.   In
our simulation we have taken the datasets of size 100 to be
subsets of those of size 1000. We fix $\Theta$  in a data-driven
way as suggested by \cite{green:rich:2001}:  if $R$ denotes the
range of the data, then  we set $\mu=R/2,\sigma_z=R,\gamma=2$ and
$\beta=0.02R^2$. Data-dependent choice of hyperparameters is
commonly made in mixture models, see for example
\cite{rich:green:mixt}.  We consider three different values of
$\alpha$,  $0.2,1$ and $5$.  We use a Gibbs move to update
$Z_j=(Z_j^{(1)},Z_j^{(2)})$ for every $j \in I^{(\al)}$: we update
$Z_j^{(2)}$ given $Z_j^{(1)}$ and the rest, and then $Z_j^{(1)}$
given the new value of $Z_j^{(2)}$ and the rest. The same scheme
is used to update the corresponding cluster parameters in the
marginal algorithms.

We monitor the convergence of four functionals of the updated
variables: the number of clusters, $M$, the deviance $D$ of the
estimated density, and $Z^{(1)}_{K_i}$, for $i=1,2$ in `lepto' and
for $i=2,3$ in `bimod'. These functionals have been used
in the comparison studies in \cite{neal:2000} and
\cite{green:rich:2001} to monitor algorithmic performance. In both cases,
the components monitored were chosen to be ones whose allocations were
particularly well- and badly-identified by the data.
The deviance $D$ is calculated as follows
$$D=-2\sum_{i=1}^n \log \big\{ \sum_{j \in
  I^{({\al})}}\frac{m_j}{n} f(Y_{i} \mid Z_j)    \big\}\,;$$
see \cite{green:rich:2001} for details. Although we have given the
expression in terms of the output of the conditional algorithm, a
similar expression exists given the output of the marginal
algorithms. The deviance is chosen as a meaningful function of
several parameters of interest.

The efficiency of the sampler is summarised by reporting for each
of the monitored variables  the estimated integrated
autocorrelation time, $\tau=1+2\sum_{j=1}^{\infty} \rho_j$, where
$\rho_j$ is the lag-$j$ autocorrelation of the monitored chain.
This is a standard way of measuring the speed of convergence of
square-integrable functions of an ergodic Markov chain
\citep{roberts-practice,soka:1997} which has also been used by
\cite{neal:2000} and \cite{green:rich:2001} in their simulation
studies. Recall that the integrated autocorrelation time is
proportional to the asymptotic variance of the ergodic average. In
particular, if $\tau_1/\tau_2=b>1$, where $\tau_i$ is the
integrated autocorrelation time of algorithm $i$ for a specific
functional, then Algorithm 1 requires roughly $b$ times as many
iterations to achieve the same Monte Carlo error as Algorithm 2,
for the estimation of the specific functional. Estimation of
$\tau$ is a notoriously difficult problem. We have followed the
guidelines in \Section 3 of \cite{soka:1997}. We estimate $\tau$
by summing estimated autocorrelations up to a fixed lag $L$, where
$\tau << L << N$, and $N$ is the Monte Carlo sample size.
Approximate standard errors of the estimate can be obtained; see
formula (3.19) of \cite{soka:1997}. For the datasets and prior
specifications we have considered we have found that $N=2\times
10^6$ suffices in order to assess the relative performance of the
competing algorithms.

 The results of our comparison are reported in  Tables
 \ref{bimod-100} - \ref{1000}. We contrast our retrospective \MCMC
 algorithm with three marginal algorithms: an improved version of the
 no-gaps algorithm of \cite{mace:mull:1998}, where we updated
 dead-cluster parameters after each allocation variable update,
 Algorithm 7 of \cite{neal:2000}, and Algorithm 8 of \cite{neal:2000},
 where we use three auxiliary states. The results show that
 Algorithms 7 and 8  perform better than the competitors,
 although the difference among the algorithms is moderate. In this
 comparison we have not taken into account the computing times
of the different methods. Our implementation, which however did
not aim at optimizing computational time, in FORTRAN 77 suggests
that no-gaps, Algorithm 7 and the retrospective \MCMC algorithm
all have roughly similar computing times when $\alpha=1$.
Algorithm 8 is more intensive than Algorithm 7. The computing time
of the  retrospective algorithm increases with the value of
$\alpha$.
%

\begin{figure}[h]
\centerline{\psfig{figure=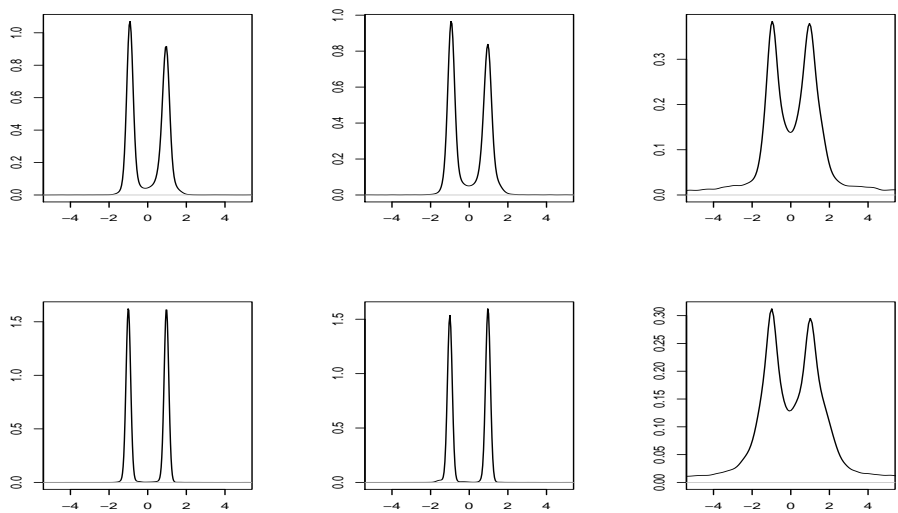,height=5in,width=3in,angle=270}}
\caption{ Posterior densities of $Z_{1}$ (left), $Z_2$ (middle)
and $Z_3$ (right) for the `bimod 100' (top) and the `bimod 1000'
(bottom) datasets. All results have been obtained for $\alpha=1$.}
\label{fig:multi2}
\end{figure}

Careful inspection of the output of the algorithms has suggested a
possible reason why the conditional approach is outperformed by
the marginal approaches. This is because it has to
explore multiple modes in the posterior distribution of the random
measure $(p,Z)$; see for example Figure \ref{fig:multi2} for
results concerning the  `bimod' dataset. On the other hand, the
ambiguity in the cluster labelling is not important in the
marginal approaches, which work with the unidentifiable allocation
structure and do not need to explore a multimodal distribution.
This indicates that the marginal approaches achieve generally
smaller integrated autocorrelation times compared to the
conditional approach. Nevertheless, the label-switching moves we
have included have substantially improved the performance of our
algorithm. 

\section{Discussion}
\label{discuss}


The appeal of the conditional approach lies in its potential for
inferring for the latent random measure, which we have
illustrated,
 and in its flexibility to
be extended to more general stick-breaking random measures than
the Dirichlet process. With respect to the latter, we have not
explicitly shown how to extend our methods to more general models,
but it should be obvious that such extensions are direct. In
particular, Proposition 1
will have to be adapted
accordingly, but all the crucial conditional independence
structure which allows retrospective sampling will be present in
the more general contexts.

In extending this work, we have already discussed in \Section
\ref{sec:acc}  an exact Gibbs sampler, i.e.\@ one in which  the
allocation variables are simulated directly from their conditional
posterior distributions. If the likelihood
function is unbounded,  this has to be carried out by an intricate
coupling of the Dirichlet process which permits tight bounds on
the normalising constants $c_i$ and also allows retrospective
simulation of all related variables. Although implementation of
 the resulting
algorithm is simple to implement, the mathematical construction
behind this method is very cimplicated and will be reported
elsewhere.

Retrospective sampling is a methodology with great potential for
other problems involving simulation and inference for stochastic
processes. One major application which has emerged since the
completion of this research, is the exact simulation and
estimation of diffusion processes
\citep{besk:papa:robe:2004,besk:papa:robe:fact,besk:papa:robe:fear:2005}.


\section*{Acknowledgement}

We are grateful to Igor Pruenster for several constructive
comments. Moreover, we would like to thank Stephen Walker, Radford
Neal, Peter Green, Steven MacEachern, two anonymous referees
and the editor for valuable suggestions.

\bibliography{dirichlet_refs}

%

%
\begin{table}
{\small
\begin{center}
\begin{tabular}{ccccc}
 & $M$ & $D$ & $Z_{K_3}$ & $Z_{K_2}$  \\
Method  &
\multicolumn{4}{c}{ $\alpha=1$}  \\
Retrospective  & 41.42 (2.6) &  3.28 (0.21)& 3.9 (0.25)&
3.44 (0.22) \\
No-gaps & 45.94 (1.52) &  3.84 (0.13) & 3.66 (0.12)    &  2.82 (0.09)  \\
Algorithm 7 & 21.85 (0.77) & 2.48 (0.09) & 3.47 (0.12)&   2.91 (0.10) \\
Algorithm 8 & 18.21 (0.66) & 2.94 (0.11) & 3.44 (0.13)& 3.1
(0.12)\\
& \multicolumn{4}{c}
{$\alpha=0.2$}  \\
Retrospective  & 67.0 (4.24) & 6.8 (0.43) & 8.01 (0.51) &
3.44 (0.22) \\
No-gaps &  39.44 (1.31) & 3.8 (0.13) & 5.93 (0.2) &  3.1 (0.10)  \\
Algorithm 7 &  24.99 (0.88)  &2.87 (0.10)   &6.32 (0.22)  &2.95 (0.10)   \\
Algorithm 8  & 22.10 (0.85)& 5.30 (0.20)&  6.8 (0.26)&3.03 (0.12) \\
& \multicolumn{4}{c}
{$\alpha=5$}  \\
Retrospective  &  21.86 (1.38) &2.82 (0.18)  &2.01 (0.13)
&2.5 (0.16)  \\
No-gaps & 57.09 (1.81) & 2.99 (0.09) & 1.67 (0.05) & 2.01 (0.06) \\
Algorithm 7 & 12.55 (0.4) &1.77 (0.06)  &1.64 (0.05)  &  1.97 (0.06)  \\
Algorithm 8 & 8.2 (0.26) & 1.77 (0.06)& 1.64 (0.05) &1.97
(0.06)
\\
\end{tabular}
\end{center}
} \caption{Estimated integrated autocorrelation times for the
number of clusters $M$, the deviance $D$, $Z_{K_3}$ and $Z_{K_2}$,
for the `bimod 100' dataset. Estimates of the
standard error in parenthesis. The initial state of all chains was
all data allocated to the same cluster with parameters drawn from
the prior. } \label{bimod-100}
\end{table}
\begin{table}
{\small
\begin{center}
\begin{tabular}{ccccc}
& $M$ & $D$ & $Z_{K_1}$ & $Z_{K_2}$  \\
Method & \multicolumn{4}{c}{$\alpha=1$}  \\
Retrospective  &  40.71 (2.58) & 31.99 (2.01) & 46.58
(2.95)& 3.04 (0.19)   \\
 No-gaps & 46.08 (1.46) & 23.93 (0.76)  & 33.19 (1.05) &2.37 (0.07)  \\  Algorithm 7 &22.98 (0.73)  &20.17 (0.64)   &28.28 (0.89)  &2.33 (0.07)   \\
 Algorithm 8  & 18.02 (0.57)    & 18.91 (0.6)& 26.71 (0.85) & 2.06 (0.07)  \\
& \multicolumn{4}{c}{ $\alpha=0.2$}  \\
Retrospective  &  239.07 (15.12) & 286.49 (18.12) &157.85
(9.99)& 14.87 (0.94)     \\
No-gaps &  127.08 (6.96) & 151.90 (8.32) &97.73 (5.35)&
7.46 (0.41)    \\  Algorithm 7 & 109.37 (5.99) &171.98
(9.42)& 86.26 (4.73)&
7.95 (0.44)     \\
Algorithm 8 & 99.06 (5.43)& 142.93 (7.83)& 82.38 (4.51)&
6.98
(0.38)\\
 & \multicolumn{4}{c}{ $\alpha=5$}  \\
 Retrospective  &13.69 (0.87) &7.38 (0.47)  & 5.9 (0.37) &
1.61 (0.1)  \\  No-gaps & 44.25 (1.4) &5.72 (0.18)
  &4.14 (0.13)  &1.36 (0.04)  \\  Algorithm 7 &  10.57 (0.33) & 5.55 (0.18) & 3.52 (0.11) & 1.33 (0.04)  \\
 Algorithm 8 & 6.32 (0.2) & 5.31 (0.17)&3.23 (0.10) & 1.29
(0.04)
\\
\end{tabular}
\end{center}
} \caption{Estimated integrated autocorrelation times for the
number of clusters $M$, the deviance $D$, $Z_{K_1}$ and $Z_{K_2}$,
for the `lepto 100' dataset. Estimates of the
standard errors in parenthesis.  The initial state of all chains was
all data allocated to the same cluster with parameters drawn from
the prior.} \label{lepto-100}
\end{table}
%
\begin{table}
{\small
\begin{center}
\begin{tabular}{cccc}
 & \multicolumn{1}{c}{`bimod 1000', $\alpha=1$} & \multicolumn{2}{c}{`lepto 1000', $\alpha=1$} \\
  & $M$ & $D$ & $M$  \\  Retrospective & 149 (7) & 254
(25) & 205 (21)
\\
 No-gaps & 91 (4) & 133 (6)  & 102 (5)
\\  Algorithm 7 &  60 (3) & 87 (4) & 99 (4)
   \\
 Algorithm 8 &  58 (3) &   112 (5) & 104 (5) \\
\end{tabular}
\end{center}
} \caption{Estimated integrated autocorrelation times for the
 `bimod 1000' and `lepto 1000' datasets. Ther results for
$D$ in the `bimod 1000' data set, $Z_{K_1},Z_{K_2}$ and $Z_{K_3}$,
for both datasets
were not markedly different across the algorithms so are omitted.}
 \label{1000}
\end{table}

\end{document}